\documentclass[a4paper]{jpconf}

\usepackage{graphicx}% Include figure files
\usepackage{dcolumn}% Align table columns on decimal point
\usepackage{bm}% bold math
\usepackage{breakurl}
\usepackage{amsmath}

\begin{document}

\title{GeV excess and phenomenological astrophysics modeling}

\author{Xiaoyuan Huang$^1$, Torsten En{\ss}lin$^2$$^3$$^4$ and  Marco Selig$^2$$^5$}
\address{$^1$Physik-Department T30d, Technische Universit\"at M\"unchen, James-Franck-Stra\ss{}e, D-85748 Garching, Germany}
\address{$^2$ Max-Planck-Institut f\"ur Astrophysik, Karl-Schwarzschildstr.
1, 85748 Garching, Germany}
\address{$^3$Ludwig-Maximilians-Universit\"at M\"unchen, Geschwister-Scholl-Platz
1, 80539 Munich, Germany}
\address{$^4$Exzellenzcluster Universe, Technische Universit\"at M\"unchen, Boltzmannstr.
2, 85748 Garching, Germany}
\address{$^5$IBM R\&D GmbH, Sch\"onaicher Stra\ss{}e 220, 71032 B\"oblingen, Germany}

\ead{xiaoyuan.huang@tum.de, ensslin@mpa-garching.mpg.de, mselig@mpa-garching.mpg.de}

\begin{abstract}
Predefined spatial templates to describe
the background of $\gamma$-ray emission from astrophysical processes, like cosmic ray interactions, are used in previous  searches for the $\gamma$-ray signatures of annihilating
galactic dark matter. In this proceeding, we investigate the GeV excess in the inner Galaxy using an alternative approach, in which the astrophysical components are identified
solely by their spectral and morphological properties. We confirm the reported GeV excess and derive related parameters for dark matter interpretation, which are consistent with previous results.
We investigate the morphology of this spectral excess as preferred by the data only. This emission component exhibits a central Galaxy cusp as expected for a dark matter annihilation signal. However, 
Galactic disk regions with a morphology of that of the hot interstellar medium also host such a spectral component.  
This points to a possible astrophysical origin of the excess and requests a more detailed understanding of astrophysical $\gamma$-ray emitting processes in the galactic
center region before definite claims about a dark matter annihilation signal can be made.
\end{abstract}

\section{Introduction}
Identifying its annihilation signatures is a promising way to probe the nature of dark matter (DM). Targets, which are favored by indirect detection of DM using $\gamma$-ray
data, should contain DM in high density, be relatively nearby, and
show little flux of astrophysical (not-DM-annihilation related) $\gamma$-rays.
The Galactic
Center (GC) region is ideal with respect to the first two conditions, however,
due to supernovae explosions injecting cosmic rays
(CRs) into the interstellar medium (ISM) and compact sources of high energy particles and radiation, it exhibits 
significant amounts of astrophysical 
$\gamma$-ray emissions \cite{vanEldik:2015qla}. Several groups have reported a spatially extended GeV $\gamma$-ray excess
from the region surrounding the GC with respect to the expected diffuse Galactic $\gamma$-ray
emission (DGE) of astrophysical origin\cite{Goodenough:2009gk,Vitale:2009hr,Hooper:2010mq,Hooper:2010im,Abazajian:2012pn,Gordon:2013vta,Huang:2013pda,Hooper:2013rwa,Daylan:2014rsa}.
It is shown that the spectral shape could be fitted by DM with mass around several tens of GeV annihilating into $b\bar{b}$ or $\tau^{+}\tau^{-}$ final states
 \cite{Calore:2014xka,Agrawal:2014oha},
and the spatial extension
of this excess could be explained by a generalized NFW profile \cite{Navarro:1995iw,Navarro:1996gj}
with an inner slope $\alpha$ = 1.2  \cite{Hooper:2013rwa,Daylan:2014rsa,Calore:2014xka}. It is considerable that the DGE model uncertainties affect the apparent
GeV $\gamma$-ray excess towards the GC, which implies considerable
systematic uncertainties for the deduced DM properties or upper limits \cite{Zhou:2014lva, Calore:2014xka, TheFermi-LAT:2015kwa}. It is also shown that $\gamma$-ray emission from unresolved millisecond
pulsars could also be the origin of the GeV excess \cite{2015arXiv150605124L,2015arXiv150605104B}.  We will present here part of the analysis performed
in \cite{Huang:2015rlu}, where we use an alternative, template-free, non-parametric, and phenomenological DGE and point source model, which significantly differs from that of the Fermi collaboration and other groups, to investigate the possible DM signal in $\gamma$-ray data. 

\section{Methods}
\label{sec:sky}
Assuming the $\gamma$-ray sky to be a superposition of
a diffuse component and a point source component, Selig et al. \cite{Selig:2014qqa} used D$^{3}$PO algorithm \cite{2015A&A...574A..74S} to decompose the observed photon
flux into these two components probabilistically while also taking  into account the instrument response and the Poisson statistics of the $\gamma$-ray events. It was shown that more than 90\% of the decomposed diffuse component at all sky locations and all investigated
energies could be further accounted for by a simple, phenomenologically constructed
two components model \cite{Selig:2014qqa}: The $\gamma$-ray spectra derived from a molecular cloud
complex in the Galactic anti-center and those derived from the southern tip of
the southern Fermi bubble \cite{Su:2010qj,Fermi-LAT:2014sfa} 
served as spectral templates in a pixel-by-pixel spectral fitting of the
nine D$^{3}$PO maps at different energies. 
Sinc the phenomenological two components description captures the dominant $\gamma$-ray properties of the Milky Way, we will take the ``cloud-like'' and ``bubble-like'' components as well as the point source model of the  D$^{3}$PO analysis by Selig et al. \cite{Selig:2014qqa} as our astrophysical Galactic $\gamma$-ray model.

Apart from these astrophysical components, we model the radial distribution of Galactic DM as a generalized NFW profile \cite{Navarro:1995iw,Navarro:1996gj} with an inner slope of $\alpha$ = 1.2. The normalization is determined by fixing the DM density at the solar radius to $\ensuremath{\rho(r_{\odot}=8.5}\ kpc\ensuremath{)}=0.4\ \mbox{GeV cm}\ensuremath{^{-3}}$. Following previous works, we investigate the most common annihilation
final states $b\bar{b}$ and $\tau^{+}\tau^{-}$ with spectrum derived from PPPC4DMID \cite{Cirelli:2010xx}.

With these assumptions and with given dark matter parameters, we can calculate the total expected $\gamma$-ray counts
\begin{equation}
\lambda^{ijk}=n_{dm}^{ijk}+\alpha_{i}n_{c}^{ijk}+\beta_{i}n_{b}^{ijk}+n_{point}^{ijk}\label{eq:lambdda}
\end{equation}
in each pixel $i$, each energy bin $j$, and for each photon detection mode $k$ (FRONT or BACK). Here $\alpha_i$ and $\beta_i$ are two free parameters to re-normalize the strength of ``cloud-like'' and ``bubble-like'' components in each pixel.  Then it is possible to compare these expected counts with the actually observed
number of photons $n_{obs}^{ijk}$ to infer these parameters $p=(m_{\mathrm{dm}},\langle\sigma v\rangle,(\alpha_i),(\beta_i))$. We do this by minimizing the objective functions given by the negative
log-likelihood 

\begin{eqnarray}
&\chi_{\mathrm{ROI}}^{2}(p) = \sum_{i\in\mathrm{ROI}}\chi_{i}^{2}(p) \nonumber\\
&\chi_{i}^{2}(p) = -2 \sum_{jk}\left[n_{\mathrm{obs}}^{ijk}\ln\lambda^{ijk}-\lambda^{ijk}-\ln\left(n_{\mathrm{obs}}^{ijk}!\right)\right]
\end{eqnarray}
for any  region of interest (ROI) pixel-by-pixel with respect to
$\alpha_{i}$ and $\beta_{i}$ while scanning through
the DM parameter subspace.

\begin{figure*}[!htb]
\centering
\includegraphics[width=0.48\textwidth]{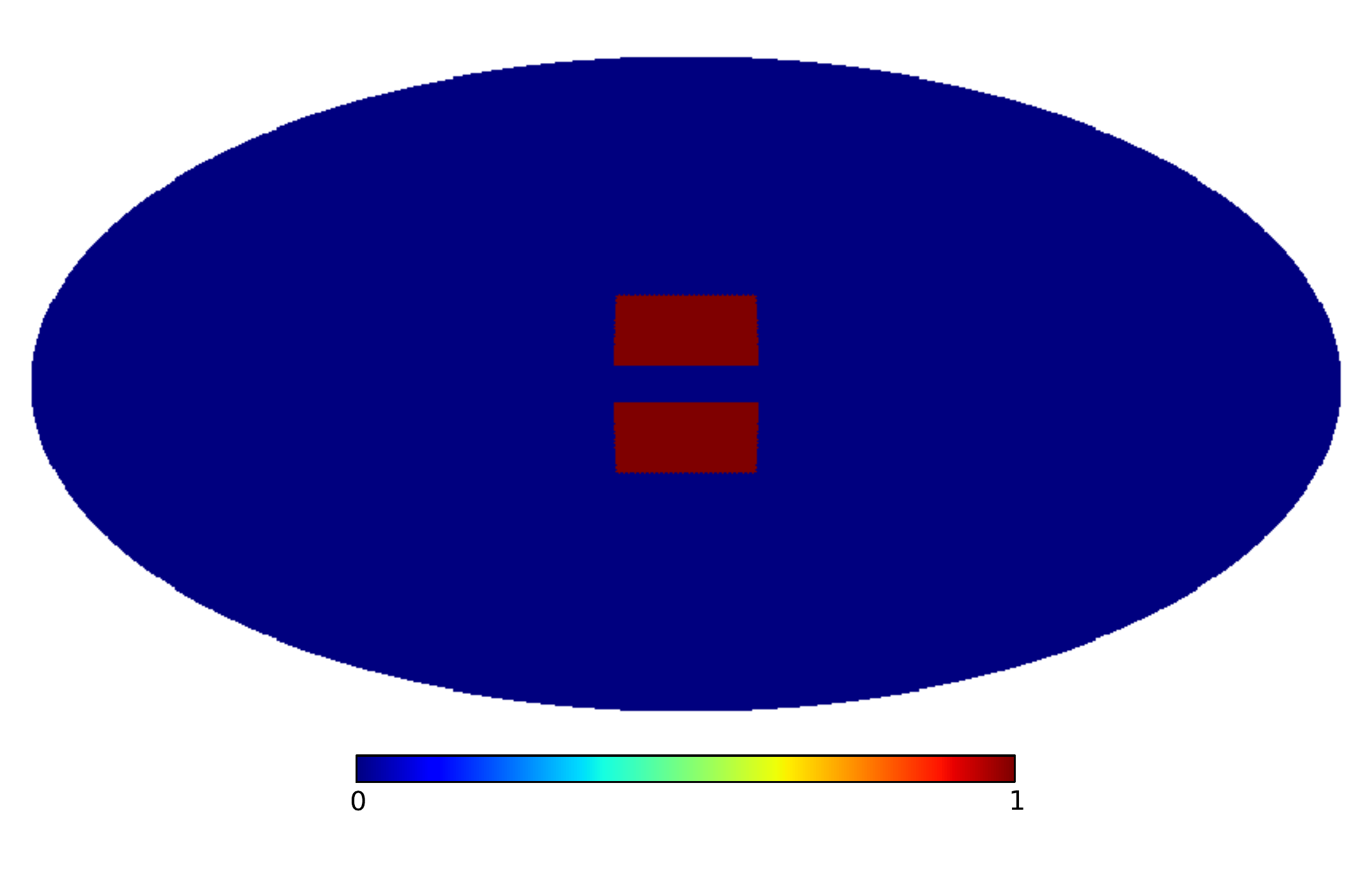}
\includegraphics[width=0.48\textwidth]{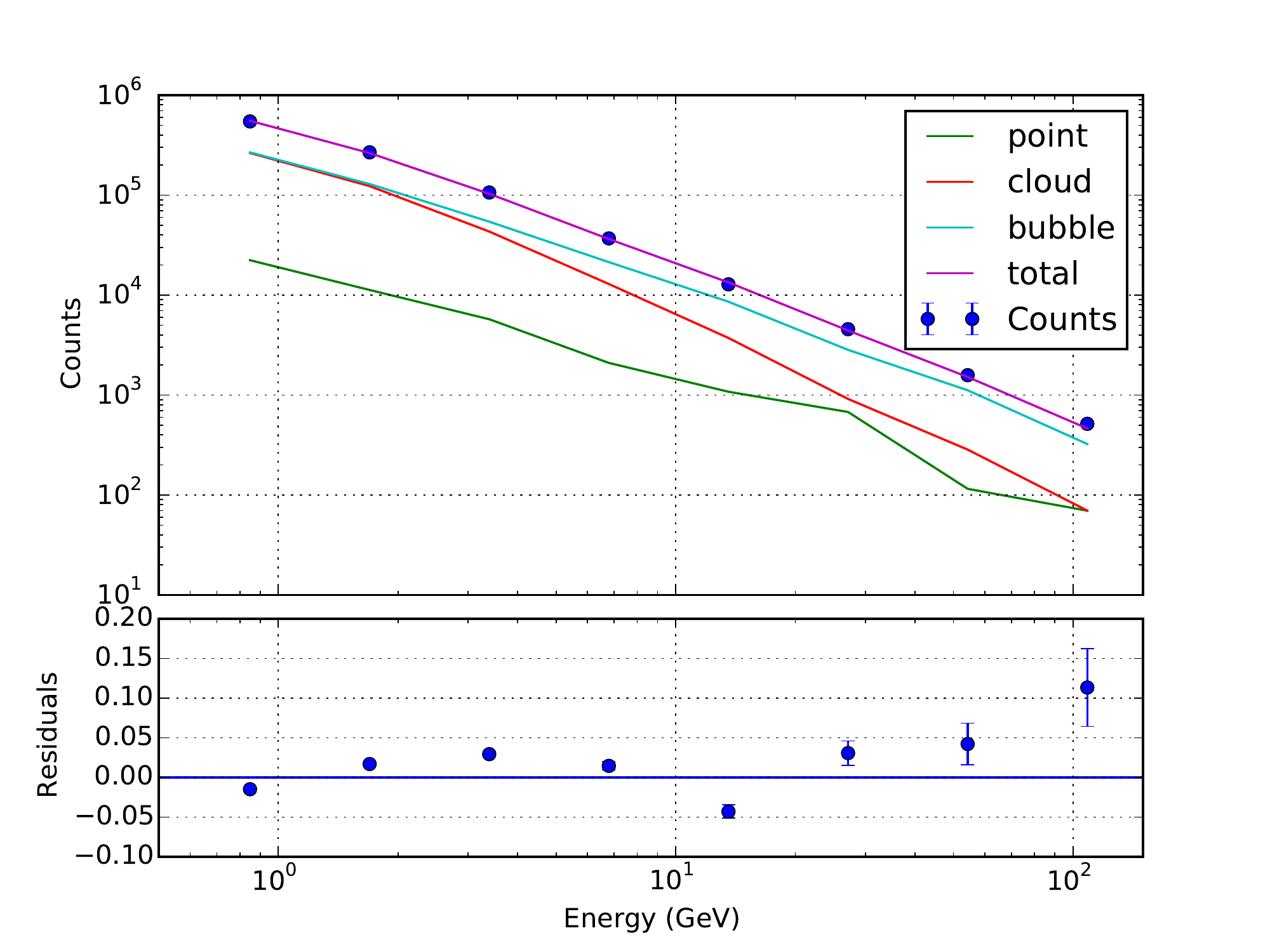}
\caption{Region used in our analysis (left) and photon counts and relative residuals for a purely astrophysical sky
model within this region (right).}
\label{fig:ROI}
\end{figure*}

Because of the complexity of the central Galactic region,
we define a ROI which excludes this area 
from our analysis. Furthermore, since
the Galactic plane
contains numerous faint, undetected and therefore not-modeled point sources, which nevertheless might contaminate
the diffuse emission, we also mask the Galactic plane for the ROI
to ensure the validity of our phenomenological two components astrophysical diffuse model. Similar to the ROI used in  \cite{Calore:2014xka}, we select Galactic latitudes $4{^\circ}<|b|<20{^\circ}$ and Galactic longitudes
$|l|<20{^\circ}$ as our ROI,
but masking a bit more of the Galactic plane region, as shown in the left panel of Fig.~\ref{fig:ROI}.

\section{Results and Discussion} 
First, we fit observational data using a purely astrophysical $\gamma$-ray
sky without DM annihilation contribution, which means to set $n_{dm}^{ijk}=0$
while fitting the remaining model parameters $\alpha_i$ and $\beta_i$
for all locations $i\in\mathrm{ROI}$. 
The right panel of Fig.~\ref{fig:ROI} shows
the observed and modeled counts within the ROI
as well as the residuals between model and data. It seems that a purely
astrophysical model fits the data reasonably well. The largest
residual appears in the bin with highest energy, where the limited
photon statistics might still cause problems to D$^{3}$PO in separating
point sources from diffuse emission. Therefore, we do not consider the
residual
at this energy as an serious indicator of DM or other new physics. 
However, around several GeV there is a small, but significant photon count
excess in the ROI. This excess seems to be coincident with the GeV excess
reported in the literature and might indicate a possible DM annihilation
signature. 

\begin{figure*}[!htb]
\centering
\includegraphics[width=0.48\textwidth]{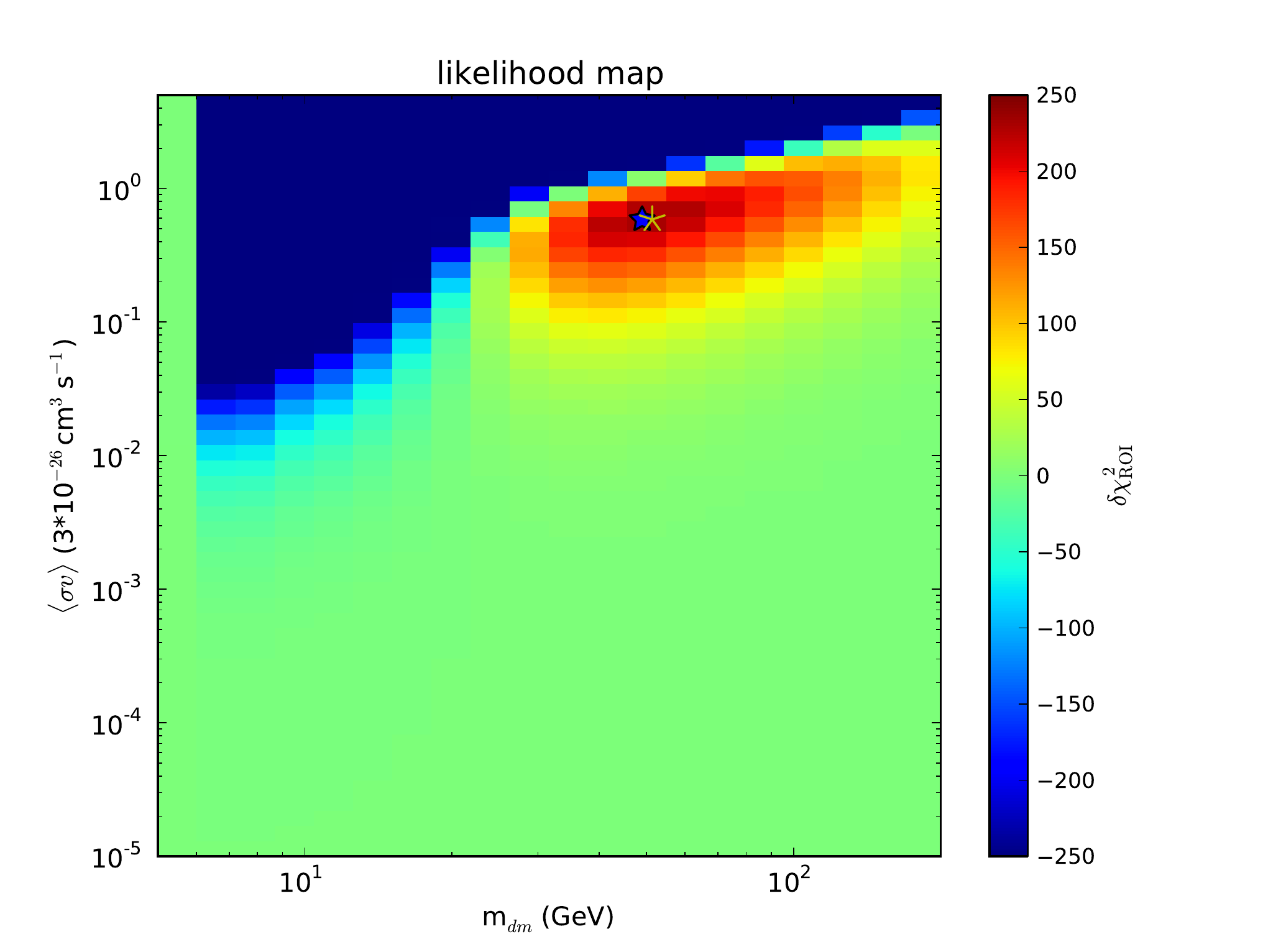}
\includegraphics[width=0.48\textwidth]{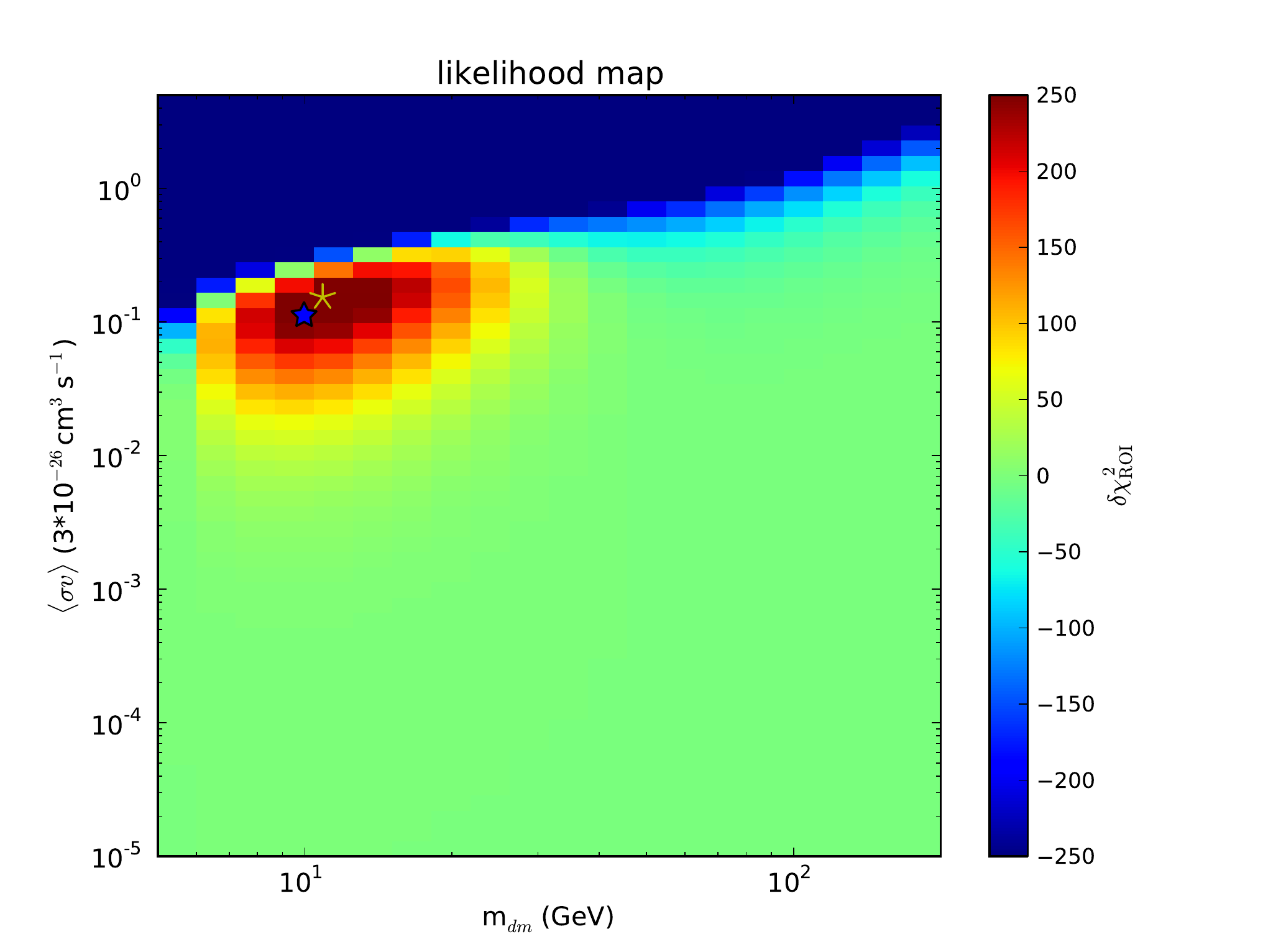}
\caption{The improvment of fitting function in our ROI for $b\bar{b}$ (left) and for $\tau^{+}\tau^{-}$
(right) annihilation final states. Best fit values from \cite{Calore:2014xka} are shown as blue stars and
our best fit DM parameters $(m_{\mathrm{dm},}\langle\sigma v\rangle)_{\star}$ are shown as yellow
stars }
\label{fig:likelihood map}
\end{figure*}

Then we scan
the dark matter
parameters $m_{\mathrm{dm}}$ and $\langle\sigma v\rangle$
while fitting the astrophysical parameter sets $\alpha$ and $\beta$ to investigate the
improvement of our objective function 
\begin{eqnarray}
 &  & \delta\chi_{\mathrm{ROI}}^{2}(m_{\mathrm{dm}},\langle\sigma v\rangle)=\nonumber \\
 &  & \mathrm{min}_{\alpha,\beta}\chi_{\mathrm{ROI}}^{2}(0,0,\alpha,\beta)-\mathrm{min}_{\alpha,\beta}\chi_{\mathrm{ROI}}^{2}(m_{\mathrm{dm}},\langle\sigma v\rangle,\alpha,\beta).
 \label{eq:ts}
\end{eqnarray}
As shown in Fig.~\ref{fig:likelihood map}, including DM with $b\bar{b}$ or $\tau^{-}\tau^{+}$
annihilation final states could indeed improve the fitting result.
The best fit DM parameters $(m_{\mathrm{dm},}\langle\sigma v\rangle)_{\star}$ agree
well with those found by Calore et al. \cite{Calore:2014xka}.

\begin{figure*}[!htb]
\centering
\includegraphics[width=0.48\textwidth]{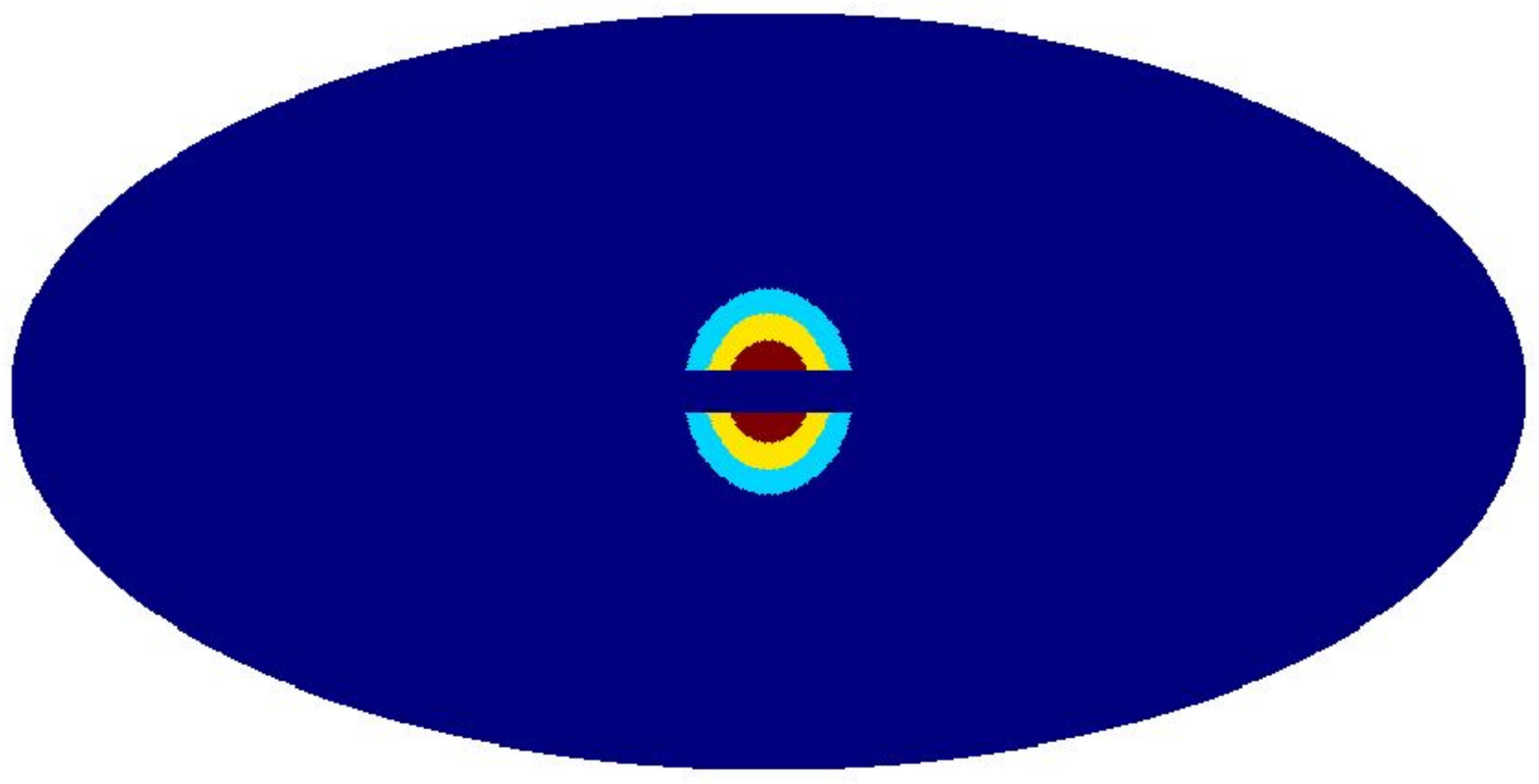}
\includegraphics[width=0.48\textwidth]{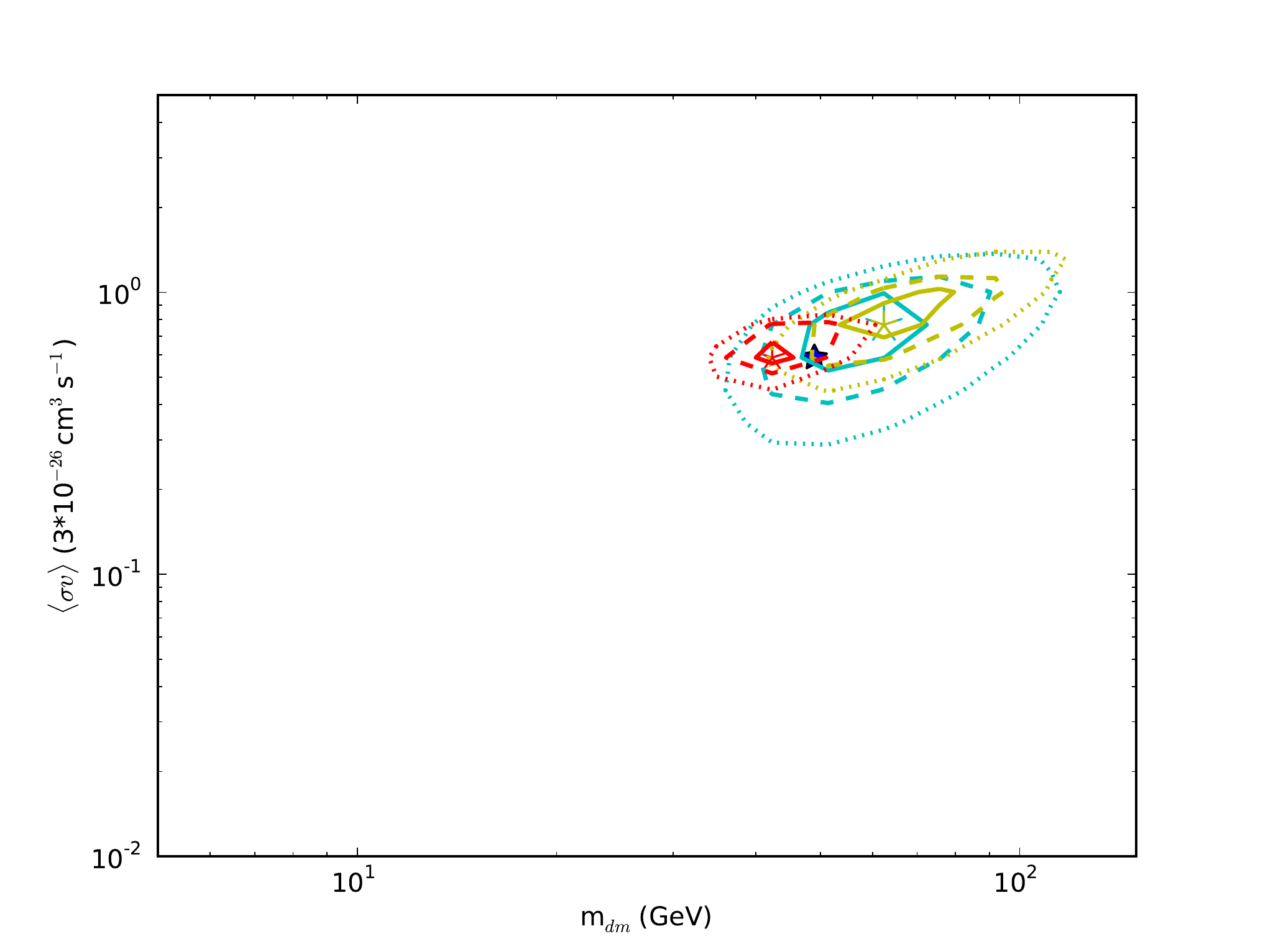}
\caption{Left: Different ROIs to test the consistency of the best fit DM
parameters $(m_{\mathrm{dm},}\langle\sigma v\rangle)_{\star}$. 
Right:  Corresponding best fit points and contours (1, 2 and 3 $\sigma$
  ) for different regions, using the same colors for the associated regions. Best fit value from \cite{Calore:2014xka} is shown as blue star.}
\label{fig:different_regions}
\end{figure*}

To verify the consistence of the best fit
DM parameters $(m_{\mathrm{dm},}\langle\sigma v\rangle)_{\star}$
inferred from different regions, we
choose three regions with different angular distances to the GC 
as in the left panel of Fig.~\ref{fig:different_regions}.
As shown in the right panel of Fig.~\ref{fig:different_regions}, derived
DM parameters are consistent with each other for $b\bar{b}$ annihilation final
states.

\begin{figure*}[!htb]
\centering
\includegraphics[width=0.48\textwidth]{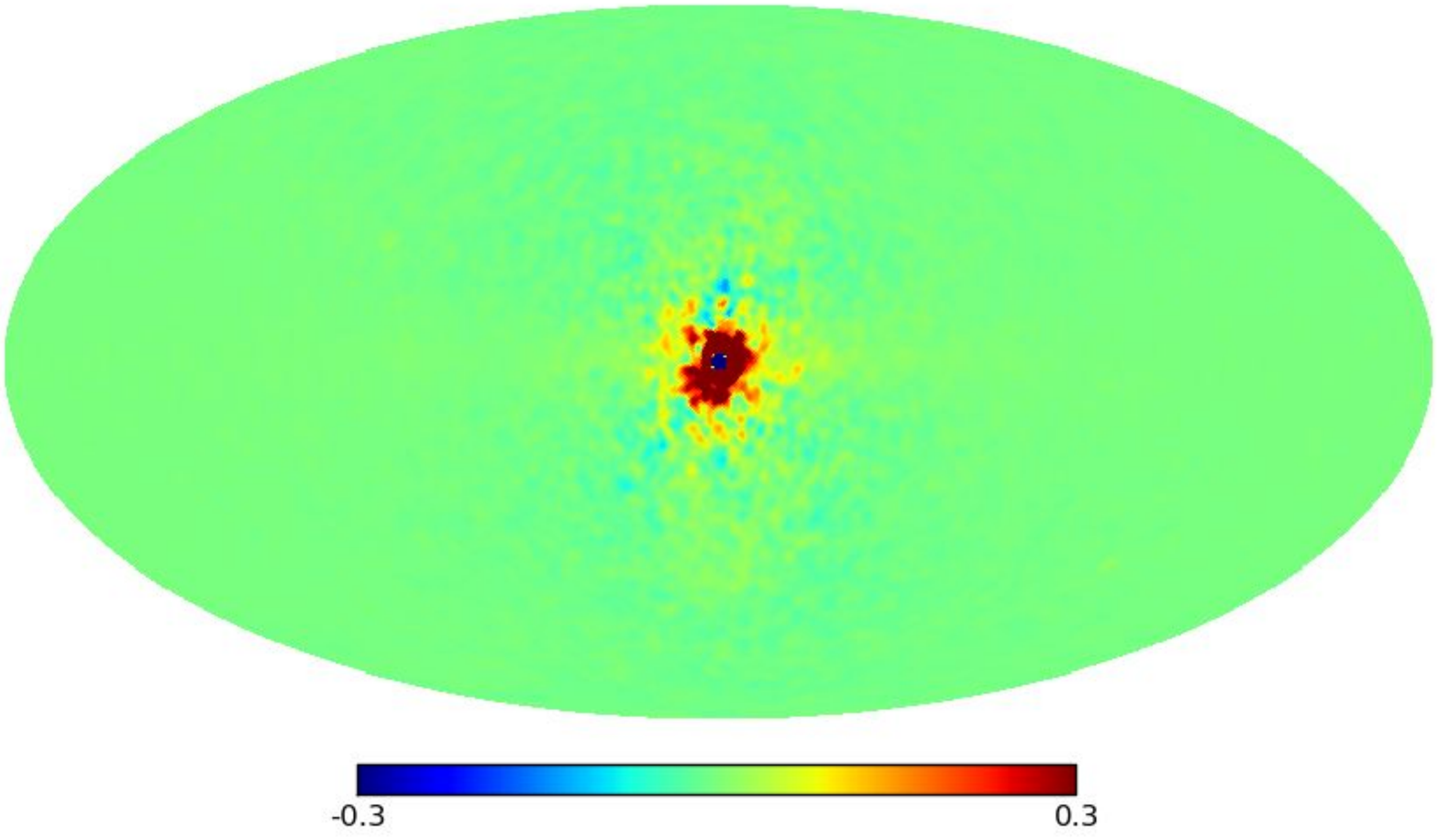}
\includegraphics[width=0.48\textwidth]{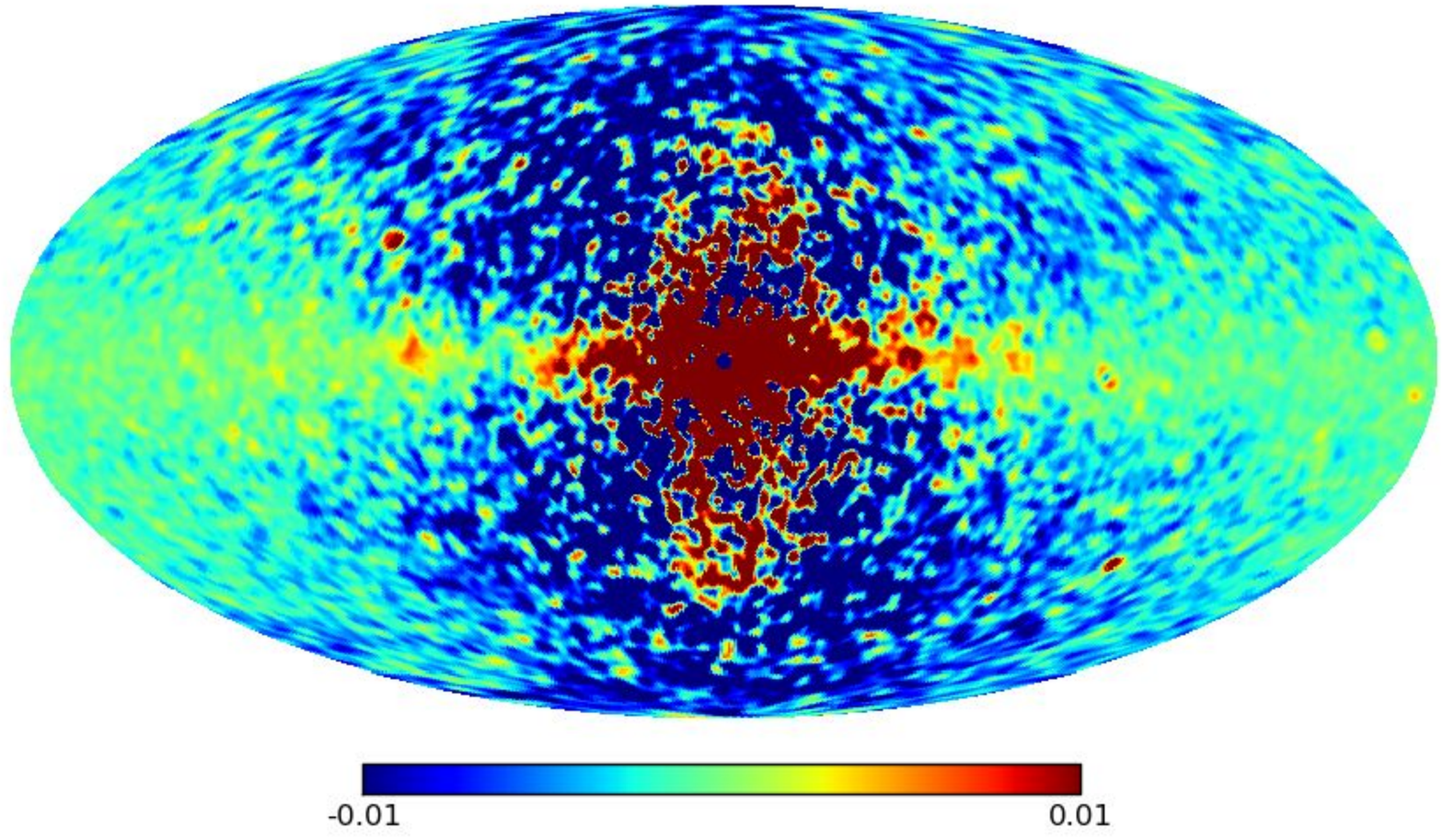}
\caption{Left: The map of the likelihood improvements $\delta\chi_{\star i}^{2}$
while including a DM component with parameters $(m_{\mathrm{dm},}\langle\sigma v\rangle)_{\star}$
for $b\bar{b}$ final annihilation states. Right: Like left panel, but with rescaled
color to highlight non-central regions.}
\label{fig:pixel_by_pixel_ts}
\end{figure*}

In order to investigate the possibility
for a potential astrophysical, non-DM annihilation
related origin of this signal, we try to find out the sky locations
driving $\delta\chi_{\mathrm{ROI}}^{2}$. To this end we use the best fit DM parameters
$(m_{\mathrm{dm}\star,}\langle\sigma v\rangle_{\star})$ to construct
all sky maps of $\delta\chi_{\star}^{2}$ as 
\begin{equation}
\delta\chi_{\star i}^{2}=\mathrm{min}_{\alpha_{i},\beta_{i}}\chi_{i}^{2}(0,0,\alpha_{i},\beta_{i})-\mathrm{min}_{\alpha_{i},\beta_{i}}\chi_{i}^{2}(m_{\mathrm{dm}\star},\langle\sigma v\rangle_{\star},\alpha_{i},\beta_{i}).\label{eq:tsi}
\end{equation}
The left panel of Fig.~\ref{fig:pixel_by_pixel_ts} shows that the improvement $\delta\chi_{\star i}^{2}$
due to the inclusion of DM annihilation contribution is almost spherically
distributed around the GC. This is consistent with the anticipation that this signal has a DM annihilation origin.
However, while we tune the colorbar (right panel of Fig.~\ref{fig:pixel_by_pixel_ts}),
the morphology of the Fermi bubbles as well as of the galactic
disk is shown at locations more distant
from the GC in in $\delta\chi_{\star i}^{2}$ map. These morphologically suspect 
regions only contribute marginally to the total $\delta\chi_{\mathrm{ROI}}^{2}$ ,
but this could indicate a problem also prevailing within our ROI since a DM contribution for the signal should
not take a morphology with astrophysical structures.
Permitting the ``DM-annihilation-like'' spectral component to exhibit any morphology preferred by the data, and not be derived from a NFW profile, we find a morphology of this component which resembles largely that of the  ``bubble-like'' component. This possibly indicates that an astrophysical spectral component in the hot interstellar medium is behind the excess emission \cite{Huang:2015rlu}.

In order to confirm or refute the apparent GeV excess as annihilation
signal, we need a better understanding of the astrophysical $\gamma$-ray radiation,
since our current sensitivity is more limited by astrophysical modeling
uncertainties than by the photon count statistics. In order to deal with
the large spatial and spectral complexity of the real Galactic $\gamma$-ray
emission, the phenomenological
and morphological methods presented here, as well as the physical
modeling approaches by other groups, need to be refined.

\ack
X.~H.~thanks the organizers of the TAUP 2015 for an interesting and
stimulating conference. This research
has been supported by the Excellence Cluster Universe.
\bigskip

\section*{References}
\bibliography{draft_gev}
\end{document}